\documentclass[aps,twocolumn,showpacs]{revtex4}
\usepackage{times}
\usepackage{graphicx}
\begin{document}
\preprint{}

\title{Shifting donor-acceptor photoluminescence in N-doped ZnO}
\author{T.~Makino}
\email[Author to whom correspondences should be addressed; electronic
mail: ]{makino@sci.u-hyogo.ac.jp}
\affiliation{Department of Material Science, University
of Hyogo, Kamigori, 678-1297, and RIKEN (Institute of Physical and 
Chemical Research), Sendai 980-0845, Japan}
\author{A.~Tsukazaki}
\author{A.~Ohtomo}
\affiliation{Institute for Materials Research, Tohoku University,
Sendai 980-8577, Japan}
\author{M.~Kawasaki}
\affiliation{Institute for Materials Research, Tohoku University,
Sendai 980-8577, Japan}
\author{H.~Koinuma}
\affiliation{Graduate School of Frontier Science, The University of
Tokyo, Kashiwa 277-8561, Japan}
%
%
\date{\today}

\begin{abstract}
We have grown nitrogen-doped ZnO films grown by two kinds of epitaxial
methods on lattice-matched ScAlMgO$_4$ substrates. We measured the photoluminescence (PL) of the two kinds of ZnO:N layers in the donor-acceptor-pair transition region. The analysis of excitation-intensity dependence of the PL peak shift with a fluctuation model has proven that our observed growth-technique dependence was explained in terms of the inhomogeneity of charged impurity distribution. It was found that the inhomogeneity in the sample prepared with the process showing better electrical property was significantly smaller in spite of the similar nitrogen concentration. The activation energy of acceptor has been evaluated to be $\approx 170$~meV, which is independent of the nitrogen concentration.
\end{abstract}
\pacs{78.55.Et, 81.15.Fg, 71.35.Cc, 72.15.-v}

\maketitle
Wide band-gap semiconductors are of intense interest for many applications,
mainly including optoelectronic devices. Many of the wide-gap semiconductors
are known to be grown more or less under residual \textit{n}-type even under
nominally undoped conditions. The trial of the \textit{p}-type doping often
results in an observation of the radiative recombination from donor-acceptor
pairs (DAPs) because of the coexistence of these two kinds of impurities.
The DAP photoluminescence (PL) bands have been widely used for more than
three decades to characterize doped semiconductors. As a result,
considerable progress has been achieved in description of the DAP bandshape
and recombination dynamics in a low-doping regime (the impurity density
being 10$^{17}$~cm$^{-3}$~\cite{tha1}). We are interested in the DAP PL for
the samples at higher doping levels (in excess of 10$^{18}$~cm$^{-3}$)
because such a large amount of nitrogen has to be doped to achieve \textit{p}-type conductivity in the case of technologically-promising wide-gap
semiconductor ZnO. Despite the tremendous world-wide activities and the
successful demonstration of homoepitaxial light emitters~\cite{nature_mat_tsukazaki}, this problem has not been solved up to now and
remains the major bottleneck of ZnO-based optoelectronics. Even when one of
the best growth techniques for ZnO:N that have ever reported [a repeated-temperature modulation (RTM) technique~\cite{nature_mat_tsukazaki}] is used,
the compensation ratio of \textit{p}-ZnO is typically as high as 0.8.

In our previous work~\cite{tamura2}, a theory
developed by Thomas, Hopfield, and Augustyniak~\cite{tha1} (THA model)
has been adopted for the analysis of the PL properties of such samples. According to this theory, the PL energy of the DAP
recombination includes the Coulomb term and is given as:
\begin{equation}
h \nu = h \nu_\infty+ e^2/\epsilon R = \left( E_G - E_A -E_D \right) + e^2/\epsilon R,
\end{equation}
where the $R $ corresponds to the inter-pair distance between donor and acceptor, while the other notations take their conventional meanings. The activation energy of acceptors ($E_A$) amounted to be as large as $\approx 260$~meV, which is much larger than the value (170~meV) obtained spectroscopically by Look and coworkers~\cite{lookp-type}, or than the value ($\approx 100$~meV) evaluated electronically by ourselves~\cite{nature_mat_tsukazaki}.

A relatively
high compensation ratio of ZnO:N films may be responsible
for a reason of this discrepancy in $E_A $. Indeed, B\"aume and coworkers
reported that their approach based on the THA theory failed to
obtain good agreement with experiments for highly-compensated ZnSe:N~\cite{baeume2}. It is thought that some of the
doped nitrogen atoms act as a donor and the total density of donors inside our doped
film is rather close to the effective density of states for donors (1--4$\times 10^{18}$~cm$^{-3}$).
In that density regime, one cannot neglect effects of overlapping of
the wave functions of somewhat delocalized donor electrons.
Because the optical transition becomes resembling with so-called ``free to acceptor'' emission, the resulting
PL spectrum is then determined by the donor-band position/width and
the position of the quasi-Fermi level.
The corresponding situation is schematically shown in Fig.~1.
On the other hand, in the THA picture, the lineshape of PL spectrum reflects
the distribution of their inter-pair distances.
Moreover, because such a high doping level tends to yield a severe inhomogeneity in a statistical distribution of involved charged impurities,
the importance of the potential fluctuation effects should be also accounted for.
\begin{figure}[htbp]
\includegraphics[width=0.45\textwidth]{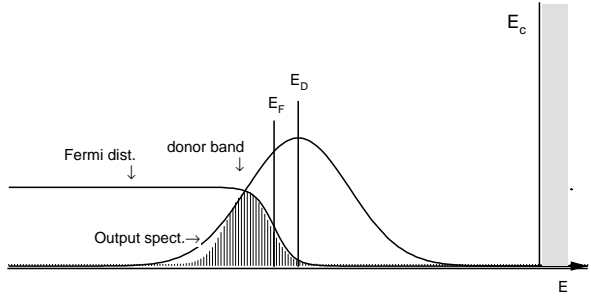}
	\caption{Contribution of the donor band part for the scheme of the spectral band formation. The hatched area corresponds to an output spectrum.}
	\label{PLrawdata}
\end{figure}

In this work, we show that, by taking the wavefunction overlapping and the
potential fluctuation effects into consideration, the earlier-mentioned
PL recombination behavior can be consistently interpreted.
\begin{table}
\caption{Sample characteristics of our ZnO:N films. The sample C was grown by the RTM technique, while the remainings were grown by the conventional method. ``HITAB'' and ``conv.'' stand for high-temperature annealed buffer and conventional growth technique, respectively.}

	\begin{center}
	\begin{ruledtabular}
	\begin{tabular*}{\hsize}{l@{\extracolsep{0ptplus1fil}}c@{\extracolsep{0ptplus1fil}}c@{\extracolsep{0ptplus1fil}}c@{\extracolsep{0ptplus1fil}}r}
	Sample&Growth temperature~$^\circ$C&[N], 10$^{18}$cm$^{-3}$&Buffer&Method\\
	\colrule
		A&650&2&---&conv.\\
		B&550&10&---&conv.\\
		C&470&7&HITAB&RTM\\
	\end{tabular*}
	\end{ruledtabular}
	\end{center} 
\end{table}

Our epilayer samples were grown on lattice-matched ScAlMgO$_4$ (SCAM) substrates by laser-beam-assisted molecular-beam epitaxy. For nitrogen doping, a radical source was operated. The characteristics of our samples have been compiled in Table~I. It is now well-known that nitrogen cannot be incorporated into
ZnO at high growth temperatures, at which the high crystallinity growth is
ensured for the undoped case. To solve this dilemma, we have developed the RTM to satisfy both high-crystallinity and high density of nitrogen. In other words, we repeated a growth sequence, in which ZnO:N layer is deposited at low-temperature, followed by rapid ramp to high-temperature, and growth of high-quality layer with smaller amount of nitrogen. The RTM epitaxy allowed one to obtain \textit{p}-ZnO when the nitrogen concentration was set to [N] $\approx 2 \times 10^{20}$~cm$^{-3}$~\cite{nature_mat_tsukazaki}.
Secondary-ion mass spectroscopy (SIMS) was used~\cite{sumiya1} for the determinations of the total N concentrations (hereafter [N]). The PL of the samples was excited by a continuous-wave
 He-Cd laser (325~nm). Maximum density of excitation was $\approx 5$~kW/cm$^{-2}$.
\begin{figure}[htbp]
\includegraphics[width=0.45\textwidth]{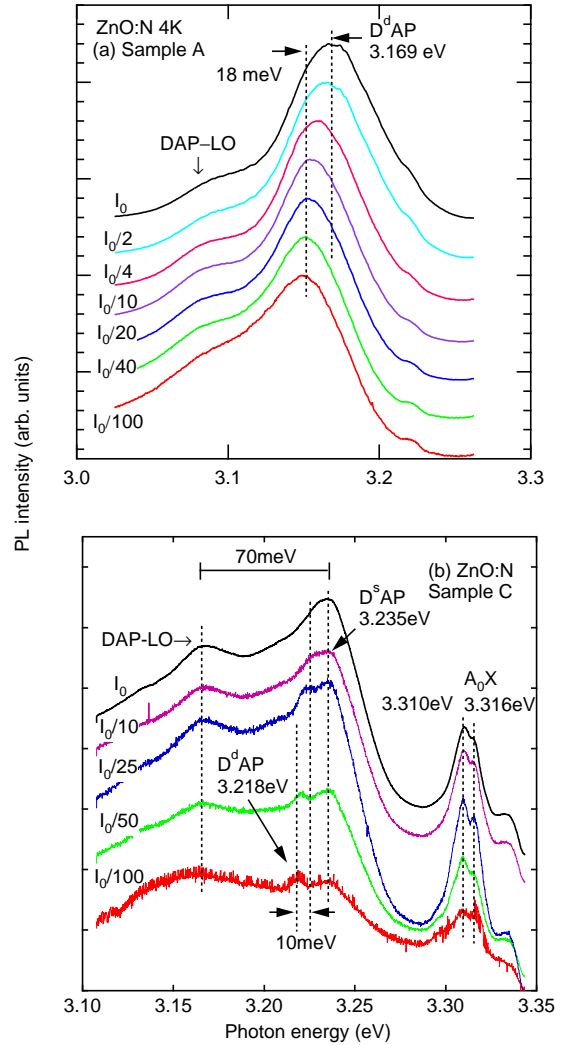}
	\caption{Excitation-intensity-dependent PL spectra taken at 5~K in nitrogen-doped ZnO films of the sample A and C. One-phonon replica (DAP-LO) is also observed for them. The emission energy of the D$^s$AP shows hardly any shift.}
	\label{PLrawdata}
\end{figure}

The evolution of low-temperature (5~K) PL spectra with the
excitation intensity for the sample A is shown in Fig.~2(a).
The sample was grown without
adopting the RTM nor inserting the annealed buffer layer.
The DAP emission band shifts to higher energies with increasing
excitation intensity. Such a shifting PL has been intuitively explained in terms of the narrowing of average inter-pair distance of the DAPs at higher laser influences.

\begin{figure}[htbp]
\includegraphics[width=0.45\textwidth]{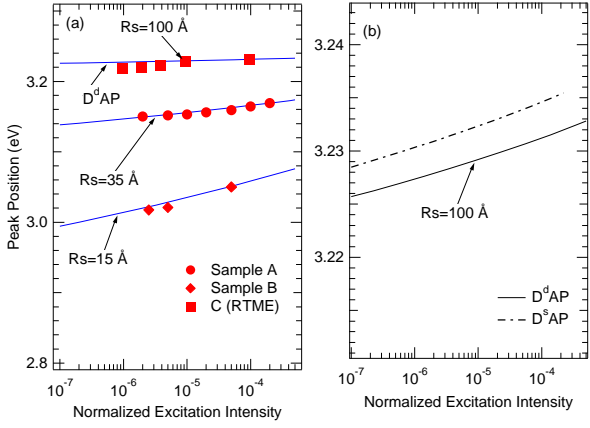}
	\caption{The dependence of peak positions on normalized excitation intensity for samples A--C. Symbols are the experimental data and solid lines are the calculated results according to Eq.~(2). The parameters used in the calculation are: $R_s = 100$~$\textrm{\AA}$ for sample C; $R_s = 35$~$\textrm{\AA}$ for sample A; and $R_s = 15$~$\textrm{\AA}$ for sample B. Also shown in subfigure (b) is the expanded scale presentation for the D$^{\rm s}$AP.}
	\label{shift}
\end{figure}
The symbols (closed circles and diamonds) of Fig.~3(a) show the excitation-intensity-dependent PL peak energies
of the DAP obtained for our ZnO:N (sample A and B). The PL peaks are excitation-intensity-sensitive in the case of samples~\cite{tamura2} A and B. The observed shift per two-decades of $I_{\textrm{exc}}$ are $\approx $18, and $\approx $45~meV. One could learn from these results that the heavier doping lead to the more
excitation-intensity-sensitive blueshifted PL.

Changing the doping and growth methods (sample C) while keeping the dopant
concentration approximately unchanged from that of `A' yielded an excitation-intensity-dependent PL
significantly
different from the above-mentioned results. The traces of Fig.~2(b) and
closed squares of Fig.~3(a) correspond to its data.
The near-band-edge spectrum [Fig.~2(b)] is dominated by the two kinds of DAP emissions at 3.218~eV and 3.235~eV. Because the energy separation of these two DAP peaks ($\approx 17$~meV) is close to the difference in the shallow- and deep-donor activation energies~\cite{look_sst}, these peaks are assigned to be shallow- and deep-donor-related DAPs (D$^s$AP and D$^d$AP). It is apparent that the D$^d$AP peak shifts in 10~meV with the two-decade increase in excitation intensity, while the D$^s$AP is insensitive to the excitation intensity.
Despite the fact
of similar density of nitrogen dopant ([N], see Table~I) with that of sample A,
the peak energies for the control samples grown by the conventional method
are relatively redshifted and sensitive to the excitation intensity change.
The growth-technique dependence of the shift is interesting to be discussed.

Recently, Kuskovsky \textit{et al.} have developed a model~\cite{kuskovsky2} to quantitatively interpret the PL of the doped materials in terms of ionic-charge distribution inhomogeneity and the wavefunction overlapping effects. As was
schematically shown in Fig.~1, this theory assumes change in the quasi-Fermi level of the donor dopants ($\mu_D$) under excitation intensity is related to the observed PL peak shift and calculates self-consistently the quasi-Fermi levels as a function of the magnitude of the fluctuations, (related to the screening radius $R_s$) and the excitation intensity. The calculation showed, in the ``fluctuation'' model, if the donor
impurity band is sufficiently narrow, the change of the quasi-Fermi level
has almost no influence on the resulting PL peak energy. It should be noted
that the effect of the moving acceptor quasi-Fermi level is much weaker because
its effective density of states is typically significantly greater.

Quantitatively, the luminescence intensity $I_E(\hat{g})$ per unit energy at an energy $E$ as a function of excitation intensity $g$ can be written as follows. We here use $E= h \nu -E_G+E_A+E_D$ as an emission energy instead of $h \nu$. It is important to realize that while $h \nu $ is always positive, $E$ can be either positive or negative, depending on the magnitude and a sign of the fluctuation energy term.

Using freezing temperature $T_g$, Eq.~(2) gives the spectrum of the zero-phonon line of DAP PL in the presence of fluctuation as a function of excitation intensity.
\begin{eqnarray}
I_E(\hat{g})=\frac{4 N_A N_D W_0 R_s^3}{\sqrt{\pi}} \sqrt{\frac{\epsilon R_s}{e^2 T_g}} \int^{\infty}_{0} du \frac{u^{5/2}}{\sqrt{u-1+\exp[-u]}} \times \cr P(u) \exp \left[ -\xi u -\eta \frac{(\tilde{E}u-1)^2}{u(u-1+\exp[-u])} \right] \times \cr \left( 1-\textrm{erf} \left[ \frac{\sqrt{-2 \ln \hat{g} u}+\sqrt{\eta} (\tilde{E}u-1)}{\sqrt{u(u+1-\exp[-u])}} \right] \right),
\end{eqnarray}
with,
\begin{equation}
P(u) = 1 [1+\tilde{n} \xi^3 u^3 \slash 6]^{-1}.
\end{equation}

Here, we used the normalized excitation intensity of $\hat{g} = -2 \sqrt{\pi} \tilde{g} \tilde{\mu_{ave}}$. The value $W_0$ is related to the transition probability of DAP and $\textrm{erf}[x]$ is the error function. It must be also explained about the other notations used in Eq.~(2): $\tilde{E}=\epsilon R_s E/e^2$; $\xi = 2 R_s /R_B$;
$\eta = e^2 \slash (4 k_{\rm B} T_{g} \epsilon R_s)$; $\tilde{n}=\pi |N_A-N_D| R_B^3$; and $\tilde{\mu}=\mu_{D} \sqrt{\epsilon R_s \slash 2 e^2 T_g}$.

Though there are three running parameters ($R_s$, $T_{\rm g}$, and $N_A-N_D$) included in this model, we keep them fixed except $R_s$ (which is intimately related to the strength of fluctuations and to the width of the donor band) for the present calculation to clear up the role of $R_s$. In all cases, we used 3.182~eV for ($E_G-E_A-E_D$), which corresponds to $E_G = 3.428$~eV, $E_A = 171$~meV~\cite{lookp-type}, and $E_D = 75$~meV~\cite{look_sst,nature_mat_tsukazaki}. It should be noted that the energy position of the bound exciton emission remained almost unchanged from the value of undoped films, probably suggesting the absence of the band-gap narrowing induced by the nitrogen doping. The solid lines in Fig.~3(a) show the calculated results based on Eq.~(2) as a function of the normalized excitation intensity whose range is $10^{-7} < \hat{g} < 10^{-3}$ for three different screening radii ($R_s$). It should be noted that the progression of $R_s$ from sample A to sample C is reasonable, i.e., smaller $R_s$ for severer redshifted emission. The emission energy ($h \nu$) becomes nearly independent of the excitation intensity if
$R_s$ becomes greater than 150~\textrm{\AA}. The calculated result for
the D$^{\rm s}$AP of Fig.~3(b) using $E_A = 171$~meV and $E_D = 60$~meV (corresponding to the shallower
donor) shows small shift, though the averaged peak energy is in reasonably good agreement with the experimental
value. It is likely that $R_s$ for D$^{\rm s}$AP is greater than that of D$^{\rm d}$AP, as large as $\sim $150~\textrm{\AA}, probably because of the difference in their densities.
Although the quantitative agreement between
theory and experiment is beyond our scope of work, all the experimental values for samples A--C could ride on positions close to the corresponding theoretical curves as shown in Fig.~3(a) in the pumping range from $\hat{g} \approx 10^{-6}$ to $\hat{g} \approx 10^{-4}$. Value of $\hat{g}$ has been evaluated by integrating a theoretical spectrum and by some averaged energy on quasi-Fermi level~\cite{kuskovsky2}. We must point out here that the effects of fluctuations ($R_s$) play relatively important role for the observed difference in the excitation-intensity-dependent shift as well as the redshifted PL energy in the samples grown by two different methods. Using $E_D = 60$~meV for the cases of samples A and B
evidently requires the reduced $R_s$ for the better matching with the experiment. But such a change results in an excitation-intensity dependence, much more drastic than the experiment.

The same conclusion in terms of inhomogeneity could be drawn by comparing the PL widths for three
samples studied in this work.
The present analysis offers alternative and complementary estimate of the fluctuations than the analysis based on the width evaluation of the DAP bands because superimposing D$^s$AP contribution tends to prevent from the precise evaluation of the broadening in D$^d$AP.

By applying the fluctuation model, we found that the effect of inhomogeneity can well explain the growth-technique dependence of the donor-acceptor pair (DAP) PL properties. The analysis allows the more precise evaluation of $E_A$ and also tells us that the improved growth technique can decrease
the potential fluctuation contained in the sample. The nitrogen atoms are likely to be regularly incorporated by faithfully following their energetics under this condition because the growth of ZnO:N can initiates on a nearly-step-free smooth surface, which presumably results in a narrower distribution of impurities. To check the reproducibility, we have also grown several RTM samples in different temperature sequences,
the dopant density of which are comprable with that of the sample C. Their
PL peak energies are hardly dependent on the exciting laser flux.

In the pioneering work on this theory, experimental data taken
for three ZnSe:N samples have been also present in which their peak shifts are
comparable with respect to each other~\cite{kuskovsky2}, which in turn
lead to: $24 \quad \textrm{\AA} \le R_s \le 35 \quad \textrm{\AA}$. It is worth while mentioning that our present
experimental work could verify that the theory is more widely applicable for
large range of fluctuations ($15 \quad \textrm{\AA} \le R_s \le 100 \quad \textrm{\AA}$). It might be interesting to compare [N] evaluated by the SIMS with
the screening radius $R_s$. These two quantities link with respect to
each other in the following formula:
\begin{equation}
N=\frac{3}{4 \pi R_s^3}.
\end{equation}
In the samples A and B, the deduced density from the above equation is
comparable with or larger than [N], while the density for the sample C
is smaller than it (probably corresponds to the density of D$^{\rm d}$
or D$^{\rm s}$).

It can be proposed that the low-temperature simple PL technique combined with this fluctuation model provides a quick tool for the assessment of the charged impurity distribution randomness and for the evaluation of $E_A$, compared to the selective PL spectroscopy or scanning capacitance microscopy, both requiring more complicated instrumentation~\cite{krtschil1}.

In summary, we have grown ZnO samples doped with nitrogen and studied their PL spectra. The RTM-grown ZnO:N sample showed both D$^s$AP and D$^d$AP bands with zero-phonon energies of 3.218 and 3.235~eV, respectively. We also discussed the PL shifting behavior and compared them with the controls grown by conventional method~\cite{tamura2}. We have explained how to extract the degree of dopant distribution inhomogeneity from the results of optical characterization. The activation energy of acceptor has been evaluated with the use of proper modeling to be $\approx 170$~meV, which is independent of the nitrogen concentration.

\begin{center}
Acknowledgements
\end{center}

The authors are thankful to Y. Segawa, N. T. Tuan, C. H. Chia, M. Sumiya, K. Tamura and Y. Takagi for helpful discussions and for providing the spectroscopic and SIMS experimental data. This work was partially supported by MEXT Grant of Creative Scientific Research 14GS0204, MEXT Grant-in-Aid under the contract No.~18760017, Iketani Science and Technology Foundation under the contract No.~0181027-A, the Asahi Glass Foundation, and the inter-university cooperative program of the IMR, Japan.


\end{document}